\documentclass[5p]{elsarticle}

\usepackage{mathtools}
\usepackage{graphicx}
\usepackage{xcolor}
\definecolor{vlgray}{RGB}{245,245,245}
\newcommand{\revis}[1]{\color{black} #1 \color{black}}

\begin{document}
\title{Wang-Landau simulations with non-flat distributions}

\author{Stefan Schnabel}
\ead{stefan.schnabel@itp.uni-leipzig.de}

\author{Wolfhard Janke}
\ead{wolfhard.janke@itp.uni-leipzig.de}

\address{Institut f\"ur Theoretische Physik, Universit\"at Leipzig, IPF 231101, 04081 Leipzig, Germany}
\date{\today}
%\author{Stefan Schnabel and Wolfhard Janke}

%\maketitle

\begin{abstract}
We show how the well-known Wang-Landau method can be modified to produce non-flat distributions. Through the choice of a suitable profile this can lead to an increase in efficiency for some systems. Examples for such an enhancement are provided.
\end{abstract}

\maketitle

\section{Introduction}

Generalized ensemble Monte Carlo methods like replica exchange \cite{REM} or the multicanonical method \cite{muca1,muca2} have been introduced some time ago to improve upon the standard Metropolis algorithm. Since then it has also been discussed how these generalized ensembles can be especially designed to increase the performance further. There are for instance strategies to select suitable temperatures for the replica exchange method \cite{PT_Katzgraber,PT_Bittner}. It is also understood \cite{Trebst,Hesselbo_ook} that the multicanonical method can be improved if one aims at a distribution in  energy that is proportional to a non-trivial profile instead of constant. Recently we have shown how simulations of spin glasses can be improved when a specially designed profile is used \cite{Fabio}. However, when it comes to the closely related and widely applied Wang-Landau method \cite{WangLandau} changing to a profile is not as straightforward as for the other methods and we are not aware of any application of such a modified Wang-Landau method. In this study we demonstrate how a non-flat Wang-Landau method can be devised through a small modification of the original algorithm and show with two examples how the performance can be improved.

\section{Wang-Landau Algorithm}
The Wang-Landau (WL) algorithm employs a function $g(E)$ to accept or reject proposed moves from microstate $\mu$ to microstate $\nu$ with probability
\begin{equation}
P_{\rm acc}(\mu,\nu)=\min\left(1,\frac{g(E_\mu)}{g(E_\nu)}\right).
\end{equation}
Here, we assume that the energy $E$ does either assume only a finite number of discrete values $E\in\{E_1,E_2,\dots\}$ or that a continuous interval $E\in[E_{\rm min},E_{\rm max}]$ is divided into a finite number of equally wide subintervals (binning) on which $g$ is constant such that there is always a finite number of values for $g$. Usually $g(E)$ is set to unity everywhere in the beginning and after each step its value for the energy of the currently occupied state $\mu_t$ is multiplied by a factor $f>1$ that is reduced over time to approach unity from above: $g'(E_{\mu_t})=f\cdot g(E_{\mu_t})$. Depending on the strategy for reducing $f$ the function $g(E)$ will converge towards or at least become very similar to the density of states $\Omega(E)$ up to a constant factor. Here we will use a different but equivalent notation. We will use a weight function $W(E)=1/g(E)$. If detailed balance holds -- which requires $f=1$ among other conditions -- it is proportional to the probability with which a microstate with energy $E$ is visited.  Since during the course of a simulation $g(E)$ and equally $W(E)$ can extend over many orders of magnitude one often stores and uses its logarithmic values and consequently
\begin{equation}
P_{\rm acc}(\mu,\nu)=\min\left(1,e^{\ln W(E_\nu)-\ln W(E_\mu)}\right)
\label{eqn:p_acc}
\end{equation}
and 
\begin{equation}
\ln W'(E_{\mu_t})=\ln W(E_{\mu_t})-\ln f.
\label{eqn:wght_upd_basic}
\end{equation}
The concept of the WL method can be expressed as follows: Reduce the probability of the energy that the walker is currently at such that over long enough periods of time states that are over-represented in the ensemble are inhibited and a quasi-steady-state is reached where the logarithmic weights $\ln W(E)$ of all energies are on average reduced equally. Then the differences between the logarithmic weights of any two energies $\ln W(E_1) - \ln W(E_2)$ and consequently the ratios of their weights $W(E_1)/W(E_2)$ remain constant even though the absolute values change. It is clear that for the standard WL method this steady state is reached if all energies are hit with equal frequency since only then the number of subtractions of $\ln f$ that each logarithmic weight $\ln W(E)$ experiences is the same for all energies. Therefore, measuring this frequency by means of a histogram $h(E)$ and testing for its flatness is a reliable way to evaluate the progress of the algorithm.

\section{Non-flat Wang-Landau Algorithm}
 The goal is now to alter the method such that the steady state is reached while the frequency is not constant but is proportional to a given profile $p(E)$. If we want to keep the basic procedure of the WL method and to change $\ln W(E)$ only at the energy of the currently occupied state $\mu_t$ it is then clear that we have to modify Eq.~(\ref{eqn:wght_upd_basic}) to
\begin{equation}
\ln W'(E_{\mu_t})=\ln W(E_{\mu_t})-\frac{\ln f}{p(E_{\mu_t})}.
\end{equation}
If now each energy is hit with a frequency proportional to the profile function $p(E)$, the accumulated changes of $\ln W(E)$ will again be independent of $E$ since $p(E)$ will cancel out. The rule for accepting updates in Eq.~(\ref{eqn:p_acc}) remains unaffected. Ignoring any error saturation, the final result of this procedure, approached when $f$ is sufficiently close to unity, is a weight function
\begin{equation}
W(E)=\frac{p(E)}{\Omega(E)}
\label{eq:non_flat_weight}
\end{equation}
where $\Omega(E)$ is again the density of states. This modification of the WL method can on the one hand be seen as the introduction of energy-dependent weight modification factors:
\begin{equation}
\tilde{f}(E)=f^{1/p(E)}.
\end{equation}
On the other hand this means that if energy-dependent weight modification factors $\tilde{f}(E)$ are used the non-flat WL algorithm will produce histograms $h(E)$ that are proportional to $1/\ln(\tilde{f}(E))\propto p(E)$.

Previously we have introduced the profile $p(E)$ as a function proportional to the desired frequency leaving it not fully determined. However, at this point its magnitude becomes relevant since it directly influences the magnitude of the changes to $\ln W(E)$. A simple choice is $p(E)\ge1$; it ensures that the weights experience changes equal to or smaller than those that would be imposed by the original WL algorithm.

One important aspect of the WL method is the way the modification factor $f$ is reduced over time. The original strategy is to sample with a constant $f$ until a sufficiently flat histogram $h(E)$ has been produced and to reduce $f$ by taking its square root. Since our stated goal is to produce non-flat distributions, flatness of the histogram is not to be expected. Now, it is the ratio of histogram and profile $h(E)/p(E)$ that will approach a constant function and can be used instead of the histogram alone.

Liang et al. \cite{Liang} as well as Belardinelli and Pereyra \cite{one_over_t} have suggested an alternative way of reducing $f$. They propose to use distinct values $f_t$ at any time $t$ and one example out of a family of possibilities for the history of $f$ is
\begin{equation}
\ln f_t= \frac{t_0}{\max(t_0,t)}
\end{equation}
for some $t_0>0$. With our modification this method can be used unchanged.

To conclude the dicussion of the modification of the WL method, we come back to the original notation with the direct approximation of the density of states 
using $g(E)$. Using Eq.~(\ref{eq:non_flat_weight}) to replace $W(E)$ by $p(E)/g(E)$ we obtain for the acceptance probability
\begin{subequations}
\begin{eqnarray}
\hspace{-0.5cm} P_{\rm acc}(\mu,\nu) &=& \min\left(1,\frac{g(E_\mu)p(E_\nu)}{g(E_\nu)p(E_\mu)}\right)\\
    &=& \min\left(1,e^{\ln g(E_\mu)-\ln g(E_\nu)}\frac{p(E_\nu)}{p(E_\mu)}\right)
\end{eqnarray}
\end{subequations}
while $g(E)$ is modified according to
\begin{equation}
\ln g'(E_{\mu_t})=\ln g(E_{\mu_t}) + \frac{\ln f}{p(E_{\mu_t})}.
\end{equation}

It should be pointed out that a variation of the WL method that allows any desired profile in $E$ instead of the constant distribution that comes with the standard WL algorithm has already been introduced \cite{Liang}. However, the proposed procedure is somewhat cumbersome since it requires the modification of the weights $W(E)$ for all values (or intervals) of $E$ at every time step \footnote{The required computational effort can be kept low by smart programming.}. The method we propose here is simpler and also more in tune with the basic principle of WL sampling.

\section{Applications}

\subsection{Proof-of-concept: Ising and Potts Models}

\begin{figure}
\begin{center}
\includegraphics[width=0.95\columnwidth]{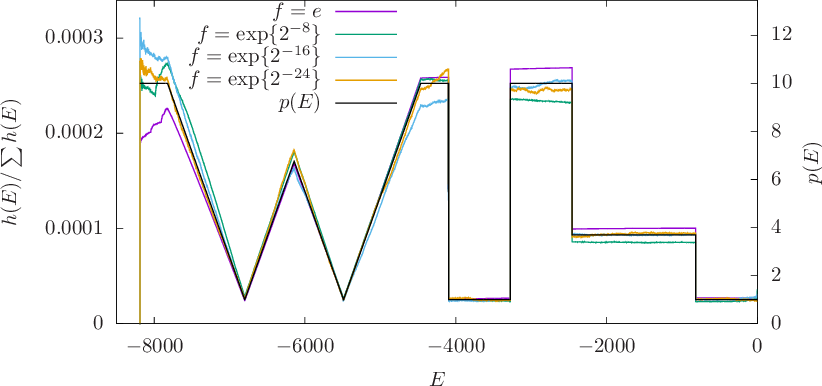}
\includegraphics[width=0.95\columnwidth]{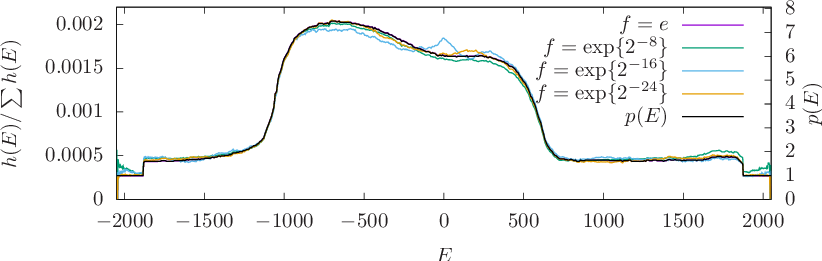}
\end{center}
\caption{\small{\label{fig:ising_potts_hist} Histograms for different values of $f$ from simulations of the $L=64,\ q=10$ Potts model with a sequence of line segments as profile (top) and the $L=32$ Ising model with a curved profile \revis{drawn to resemble a figure in} \cite{deStExupery} (bottom) on square lattices.}}
\end{figure}

In order to demonstrate that the method is working, i.e., that it is able to produce histograms in accordance with desired profiles we performed as basic test simulations of a $L=32$ Ising model and a $L=64,\, q=10$ Potts model on square lattices with periodic boundary conditions. For the reduction of $f$ we modified the original recipe from \cite{WangLandau} for our algorithm through replacing $h(E)$ by $h(E)/p(E)$: The modification factor is reduced according to $f'=f^{1/2}$ and the histogram reset to $h(E)\coloneqq0$ if the minimum of $h(E)/p(E)$ is larger than a certain fraction of its mean
\begin{equation}
\min\left(h(E_i)/p(E_i)\right)\ge\frac\rho B\sum\limits_{i=1}^B h(E_i)/p(E_i),
\label{eqn:flat_hist}
\end{equation}
where $B$ is the number of values (subintervals) of the energy and we chose $\rho=0.9$. The results displayed in Fig.~\ref{fig:ising_potts_hist} show that the method works as expected and that the histograms reproduce the desired profiles very well.

\subsection{Ising Spin Glass}
Next, we consider an Edwards-Anderson spin glass \cite{EA} on a cubic lattice with the Ising Hamiltonian
\begin{equation}
\mathcal{H}=-\sum\limits_{\langle ij\rangle}J_{ij}s_is_j,\qquad s_i,J_{ij}\in\{-1,1\},
\end{equation}
where the sum goes over all pairs of adjacent spins and the bonds are randomly chosen for any new disorder realization.
\begin{figure}
\begin{center}
\includegraphics[width=0.95\columnwidth]{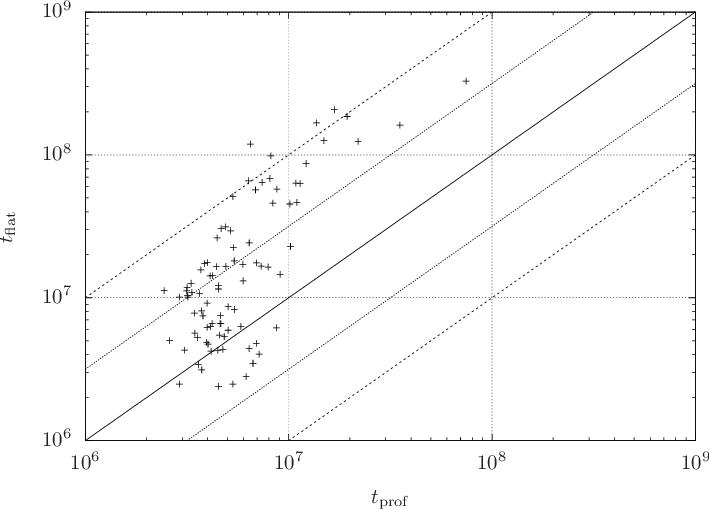}
\end{center}
\caption{\small{\label{fig:spg_times} Scatter plot of the simulation times required for individual spin-glass samples.}}
\end{figure}
Due to their rough energy landscape spin-glass systems pose an interesting challenge and serve as benchmark cases for Monte Carlo methods. In a recent study \cite{Fabio} it has been demonstrated that for equilibrium ($f=1$) simulations a power-law profile with a strong emphasis of low-energy states is superior to a flat histogram and allows a more rapid exploration of state space. We expect that a similar acceleration can be achieved for $f$ close to but larger than unity and that the performance of the WL algorithm can thus be improved. Note that the experiment described here is kept rather simple. It is intended to be a proof-of-concept and further improvements to the method are possible. We generate 100 disorder realizations $\{J_{ij}\}$ with $N=10^3$ spins and run WL simulations for all of them with a constant profile $p(E)=1$ as well as with
\revis{
\begin{equation}
p(E)=\left(\frac{E}{1896}+1\right)^{-3},
\label{eq:power_law_prof}
\end{equation}
}
which is based on the profile that was used for $N\le8^3$ in \cite{Fabio}. \revis{The exponent was changed from $-3.6$ to $-3.0$ to allow the sampling of the wider energy interval needed for the larger systems.} In the beginning $f=e$ and the simulation is stopped if $f<\exp\left\{10^{-8}\right\}$.
%After an initial phase of $10^6N$ steps every $10^4N$ steps it is tested whether all energies that had been found previously have been visited again during the current iteration, i.e., with the present $f$. If this is found to be the case the histogram is reset to zero and f is reduced $f'=f^{1/\sqrt{2}}$. The idea is to require the equivalent of minimal flatness, but reduce $f$ only half as fast.
After an initial phase of $10^6N$ attempted spin flips (steps) every $10^4N$ steps it is tested whether according to Eq.~(\ref{eqn:flat_hist}) a flat histogram with $\rho=0.2$ has been produced on the energies that have been visited during the current and previous iterations. If this is found to be the case the histogram is reset to zero and $f$ is reduced: $f'=f^{1/2}$.
If during the simulation a new lowest energy is found and if $f<\exp\left\{10^{-5}\right\}$ we reset $f=\exp\left\{10^{-5}\right\}$ to allow the simulation to adjust the weight of the states at the new energy and prevent it from getting trapped there.

The required Monte Carlo time in units of $N$ steps for \revis{89} disorder realizations is shown in Fig.~\ref{fig:spg_times}. For the remaining \revis{11} samples the two methods did not find the same lowest energy and can, therefore, not be compared. In \revis{ten} cases the WL simulation with the power-law profile reached a lower energy while the flat-histogram version reached a lower energy once. Where comparisons \revis{of running time} are possible we see that for the hard samples, i.e., the disorder realizations that require long simulations, the power-law profile in Eq.~(\ref{eq:power_law_prof}) is more than \revis{four} times faster while it \revis{can lead} to less efficient simulations for the very easy samples. The aggregated running times for all \revis{89} samples shown are \revis{$6.42\times10^{8}$} for the power-law profile vs \revis{$2.96\times10^{9}$} for the flat distribution.

\subsection{Lennard-Jones Polymer}
\begin{figure}
\begin{center}
\includegraphics[width=0.75\columnwidth]{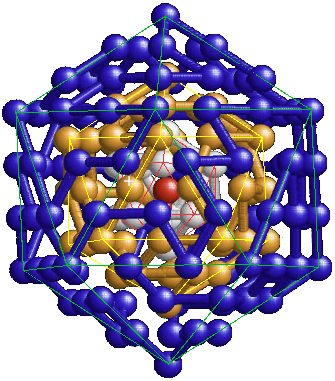}
\end{center}
\caption{\small{\label{fig:emin147} Icosahedral low-energy conformation of the polymer. Colors indicate the different layers containing 1,12,42, and 92 beads.}}
\end{figure}

As a second example for a useful application we apply the method to a Lennard-Jones polymer. We investigated this system some time ago and details of the model and the results can be found in \cite{LJ1,LJ2,LJ3,LJ4}. For our current purpose it is sufficient to say that it is an off-lattice bead-spring polymer model that for the considered size of $N=147$ beads possesses a very stable state of icosahedral geometry (Fig.~\ref{fig:emin147}) at low temperatures.
At medium temperatures one observes an unstructured dense globular droplet and at high temperatures beyond the so-called $\Theta$-point we find extended conformations that resemble self-avoiding random walks. Therefore, there are two transitions \footnote{These should not be considered phase transitions in the strict sense since we deal with a finite system.}, one at energies around $E\approx-670$ and one at $E\approx-170$. The former is first-order-like and poses a substantial obstacle to the walker due to the high free-energy barrier associated with it. We use a very simple profile that is defined by
\begin{equation}
p(E)=\left\{
\begin{array}{lcr}
\Pi,& {\rm if}& |E+670|\le 10\\
1,& {\rm else},&
\end{array} \right.
\label{eq:box_prof}
\end{equation}
in order to enhance the simulation. We perform WL simulations for different values of $\Pi$. We require minimal flatness of the histogram, i.e., we proceed to the next iteration when all energies have been visited at least once. Then the modification factor is reduced according to $f'=f^{1/2}$ until $\ln f\le10^{-9}$(starting with $f=e$). For the Monte Carlo updates we use an elaborate set of moves that is discussed in detail in \cite{LJ_moves}. 

At extreme low energies the system undergoes a final energy optimization which does not significantly affect the position of the beads but rearranges the bonds such that unfavorable distances are avoided if possible. This has for instance the effect that at the global energy minimum conformation -- the lowest microstate we found with $E=-805.161$ is shown in Fig.~\ref{fig:emin147} -- only exactly one bond connects the different layers. This process slows down the simulation in the proximity of the ground state and one might try to counter this by an additional increase of the profile in this region. However, here we just want to look on the effect of the `solid-liquid' transition and hence exclude the ground state by restricting the energy range to $-800\le E\le200$. 

\begin{figure}[t]
\begin{center}
\includegraphics[width=0.99\columnwidth]{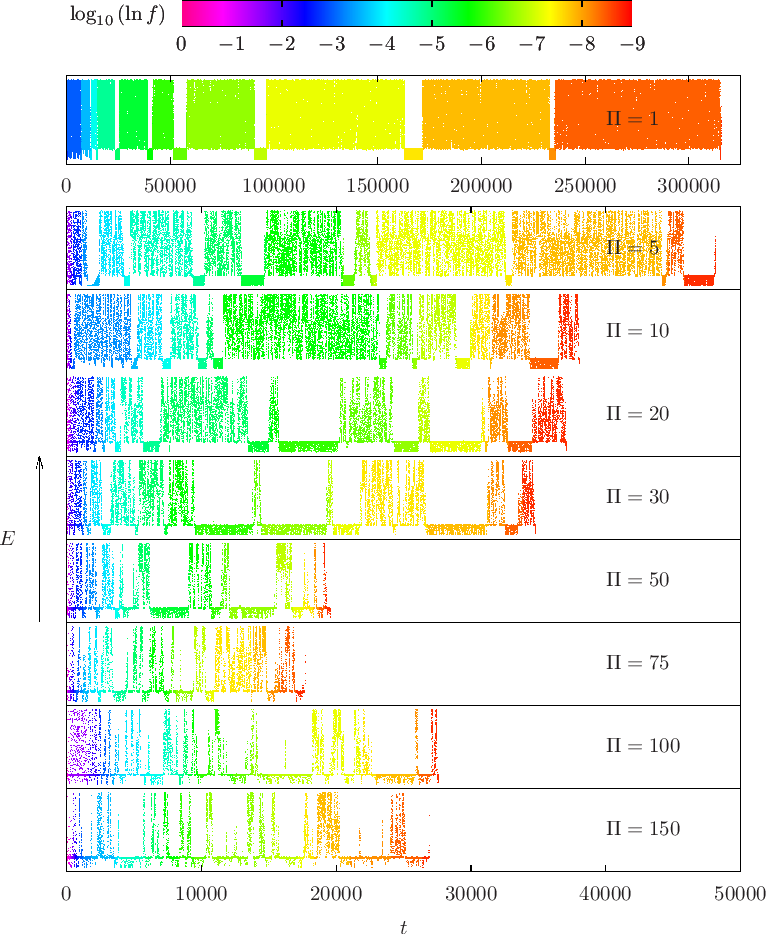}
\end{center}
\caption{\small{\label{fig:fene_ts} Time series for different levels of enhancement $\Pi$ of the transition region using the profile (\ref{eq:box_prof}). All time series cover the same vertical energy range $E\in[-800,200]$. The Monte Carlo time $t$ is measured in units of $10^3N=1.47\times10^5$ updates which is also the interval between individual points in the plots. \revis{Note the different time scales for $\Pi=1$ and $\Pi>1$.}}}
\end{figure}

Individual time series, i.e., the polymer's energy as function of time, for the different values of $\Pi$ are shown in Fig.~\ref{fig:fene_ts}. As expected the `speed' of the simulation is mainly depending on the frequency of the transitions between the `liquid' and the `solid' state at low $E$. \revis{Raising the profile $p(E)$ in the transition region} enhances this frequency and thus accelerates the simulation. \revis{However, if too much statistical weight is concentrated at the low-energy transition, the proper sampling of some other region(s) becomes the bottleneck and the simulation length increases again.} Although the length of individual WL simulations is to some extent subject to chance and changes with the seed of the random number generator, the general trend is obvious and $\Pi\in[50,75]$ is \revis{ about fifteen times faster than a flat ($\Pi=1$) distribution and also superior to an even more distorted ($\Pi\ge 100$) profile.}

This method should work in all cases where such a well localized single bottleneck is hampering the random walker and we expect that also studies of other systems with first-order-like phase transitions can benefit from it.

\revis{
\section{Chosing the Profile}

Although our technique of using WL with a profile has now been presented, the elephant in
% Dear lector, this is a silly little joke. It would be great if it could be kept, but feel free to remove it if necessary. (No need to get back to us)
the} \hspace{-3.5pt}\color{vlgray}b\revis{room\color{vlgray}i}\hspace{-2.5pt}\revis{ remains:
%the room remains:
How to select a suitable profile? Unfortunately, this is not a question that at the current time can be answered to complete satisfaction. Several characteristics of the stochastic process that is to be optimized have to play a role in the selection of the profile such as diffusivity depending on energy, critical slowing down in the proximity of phase transitions, or walks in rough energy landscape with a multitude of metastable states for glassy systems, and a comprehensive strategy incorporating all these factors is lacking. However, this does not mean that one could not in some cases apply heuristic methods with considerable success. For systems with strong phase transitions like the Lennard-Jones polymer discussed earlier the transition region can be identified in the simulation's early stages, i.e., for larger $f$ and the profile can be applied for the remainder still providing a substantial benefit. If a large number of similar systems have to be investigated as is typically the case for spin glasses a small subset can be used for multiple simulations with several candidate profiles thus establishing a suitable profile in a trial-and-error fashion.
}

\section{Conclusion}

We have shown how the Wang-Landau method can with minimal effort be adapted to produce non-flat histograms with a desired profile for any value of the modification factor.
As expected the advantages of balanced simulations with a profile \cite{Fabio} carry over to Wang-Landau sampling. Spin glasses can be simulated more efficiently with the here proposed non-flat Wang-Landau algorithm if the profile is high at low energies and it is likely that similar gains can be achieved for other glassy systems as well. As shown in the case of a polymer, transitions between different macrostates occur more often if the profile is enhanced in the transition region and the performance of the simulation method can thus be increased.

To introduce the concept of our non-flat Wang-Landau method, we have in this \revis{article} focused on an implementation based on Monte Carlo simulations as in the original publication \cite{WangLandau}. It is, however,  also easily possible and straightforward to boost standard flat Wang-Landau molecular dynamics simulations \cite{WL_MD_1,WL_MD_2} by employing a non-flat generalization along the same lines as discussed here.

\revis{For now, this technique is no more but also no less than another item in the toolkit of Monte Carlo methods. We hope its introduction will encourage further research into efficient and reliable ways to obtain useful profiles for broad-histogram simulations so that more comprehensive strategies allowing for more powerful algorithms might be conceived.}

\section*{Acknowledgement}
This  project  was  funded  by  the  Deutsche  Forschungsgemeinschaft  (DFG,  German  Research  Foundation)  under Project No.\ 189 853 844–SFB/TRR 102 (project B04). It was further supported by the Deutsch-Französische Hoch\-schule (DFH-UFA) through the Doctoral College ``L4'' under Grant No. CDFA-02-07.

\end{document}